\documentclass[prl,aps,twocolumn,amsmath,amssymb]{revtex4}
\usepackage{graphicx}
\usepackage{amsmath}
\usepackage{amssymb}
\begin{document}

\title{Transmission Phase Through Two Quantum Dots Embedded in a Four-Terminal Quantum Ring}

\author{M. Sigrist,$^1$,A. Fuhrer,$^1$ T. Ihn,$^1$ K. Ensslin,$^1$ W. Wegscheider,$^2$ M. Bichler$^3$}

\affiliation{
$^1$Solid State Physics Laboratory, ETH Z\"urich, 8093 Z\"urich, Switzerland\\
$^2$Institut f\"ur experimentelle und angewandte Physik, Universit\"at Regensburg, Germany\\
$^3$Walter Schottky Institut, Technische Universit\"at M\"unchen, Germany}

\date{\today}

\begin{abstract}

We use the Aharonov-Bohm effect in a four-terminal ring based on a Ga[Al]As heterostructure for the measurement of the relative transmission phase. In each of the two interfering paths we induce a quantum dot. The number of electrons in the two dots can be controlled independently. The transmission phase is measured as electrons are added to or taken away from the individual quantum dots. 
\end{abstract}

\maketitle


Two-terminal interference measurements in na\-no\-struc\-tures do not allow the determination of the relative transmission phase due to the generalized Onsager relations~\cite{86Bbuttiker}.
In a series of recent experiments it has been demonstrated~\cite{97schuster, 98buks,00ji,02ji} that in a multi-terminal geo\-metry the relative transmission phase can be directly observed. In the quantum rings used in these expe\-riments, two spatially separate transmission channels interfere and the phase difference of the corresponding transmission amplitudes can be detected by measuring Aharonov-Bohm (AB) oscillations.
By using one of the two interfering paths as a reference the phase evolution in the other path can be studied, e.g. when a 
Coulomb blockaded quantum dot is embedded there~\cite{97schuster}.
It was found that the transmission phase changes by about $\pi$ across a typical Coulomb blockade resonance and regularly exhibits so-called phase lapses between resonances.
Similar experiments were carried out on a Kondo-correlated system in Refs.~\cite{00ji} and~\cite{02ji}. 
A large number of theoretical papers (for a review see~\cite{01hackenbroich}) has addressed the issue of the phase lapses.

Here we investigate the phase evolution of a system of
two quantum dots with negligible electrostatic interaction embedded in two arms of a
four-terminal AB ring. Both arms of the ring including the two dots can be tuned individually. As a single electron is added to either of the
two quantum dots by increasing the corresponding plunger gate voltage, the observed average phase shift is about $\pi$ but smaller shifts are also observed. 
In such a measurement one dot is kept on a conductance resonance while the
phase evolution induced by tuning the other dot through a resonance is
moni\-tored.  In contrast, if a single electron is added to each of the quantum dots simultaneously, the observed phase
shift is close to zero, since the relative phase change in each arm is roughly the same.

\begin{figure}
\centering
\includegraphics[width=7.5cm]{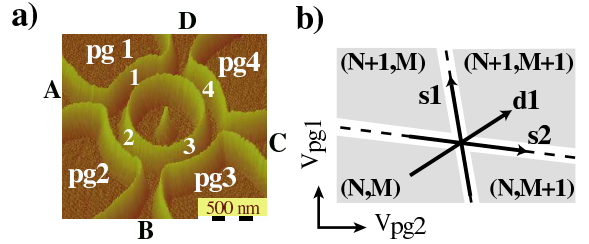}
\caption{\label{fig1} (a) AFM micrograph
of the ring structure. The oxide lines (bright lines) fabricated by AFM
lithography lead to insulating barriers in the two-dimensional electron gas. The areas marked pg1-pg4 are used as lateral gates to tune the conductance of the four arms of the ring and act as
plunger gates for the dots. The four terminals of the ring are labeled A through D. (b) Schematic plot of the parameter space defined by $V_{pg1}$ and $V_{pg2}$. The dashed lines mark the conductance maximum positions. AB-oscillations were measured along the arrows. N(M) denotes the electron occupation in dot 1(2) respectively.}
\end{figure}

The sample is a Ga[Al]As heterostructure containing a
two-dimensional electron gas 37 nm below the surface. The lateral
pattern was fabricated with the biased tip of an atomic force
microscope which locally oxidizes the GaAs surface. Details of this fabrication technique are described in
Ref.~\cite{02fuhrer}. Figure~\ref{fig1}(a) shows an AFM micrograph of
the oxidized pattern.  Lateral gate electrodes labeled pg1
through pg4 are used to tune the conductance in each of the four
segments of the ring. 
All measurements were carried out at 100\,mK in a dilution refrigerator.

\begin{figure}
\centering
\includegraphics[width=7cm]{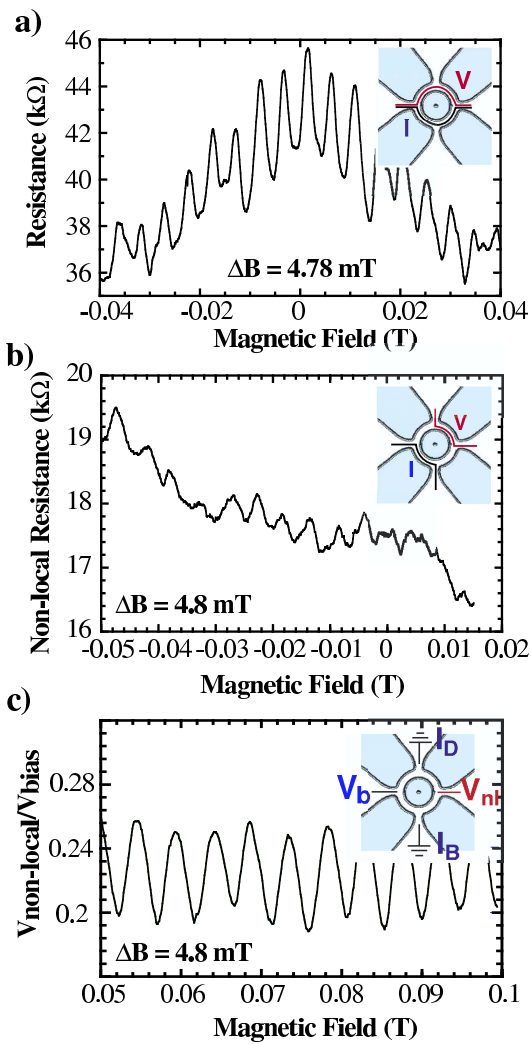}
\caption{\label{fig11} (a) AB-Oscillations in a local measurement setup (see inset). (b+c) Two different non-local setups. The measurement setup shown in the inset of (c) turned out to show the largest AB-Oscillations and was therefore used for the phase sensitive measurements.}
\end{figure}

The ring was characterized in the open
regime where each segment supports 2-4 lateral modes. Figure~\ref{fig11}(a) shows a
pronounced AB effect in a local measurement setup
in which the current and voltage contacts are the same two terminals
of the ring [inset Fig.~\ref{fig11}(a)]. The period  $\Delta B=\Phi_0 /A\approx4.8$~mT is in agreement with the area $A=0.85$~$\mu$m$^2$ enclosed by the ring.

The non-local resistance was measured by passing a current through arm 2 and measuring the voltage drop across arm 4. This is shown in Fig.~\ref{fig11}(b). In this case the AB-oscillations constitute only 2\% of the total signal.  In order to maximize the AB signal we therefore chose a different arrangement [inset Fig.~\ref{fig11}(c)] similar to the one used in Ref.~\cite{97schuster}. Here a bias voltage $V_\mathrm{b}$ was applied to terminal A. The lower and
upper contacts (B and D) were grounded via current-voltage
convert\-ers measuring the currents $I_{\mbox{\tiny B}}$ and $I_{\mbox{\tiny D}}$. The non-local voltage $V_\mathrm{nl}$ was measured at terminal C. The amplitude of the AB-oscillations in $V_\mathrm{nl}$ [Fig.~\ref{fig11}(c)] was found to be about 25\% of the total signal.

\begin{figure}
\centering
\includegraphics[width=7.5cm]{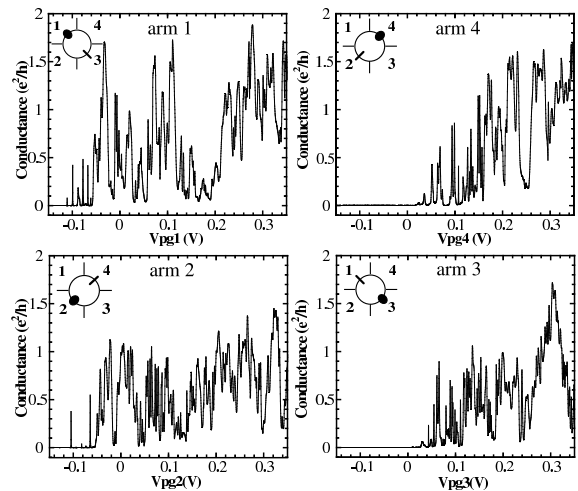}
\caption{\label{fig2} Coulomb blockade oscillations in the conductance through each arm of the ring measured in a two-terminal configuration. For each measured arm the opposite arm was pinched off by applying a strongly negative voltage to the corresponding gate. The schematic at the top left in each figure depicts the measurement setup where the line marks the arm which is pinched off and the black dot indicates the arm which is tuned.}
\end{figure}

We also characterized the four arms of the ring in the closed regime. To this end we measured the two-terminal conductance of each segment as a function of gate voltage when the arm on the opposite side was pinched-off by applying a strongly negative voltage to the corresponding plunger gate. In Fig.~\ref{fig2} the resulting conductance sweeps are shown. For low enough gate voltages clear Coulomb blockade oscillations are observed indicating the formation of a quantum dot in the arm tuned. 

The location of the quantum dots can be estimated from the respective lever arms to each of the gates pg1 to pg4 which reveals that the dots form within the segments and not at the openings to the contacts. The lever arm for the gate closest to the corresponding dot is found to be $\alpha_{\mathrm{G}}\approx0.16$, for the two adjacent gates the lever arm is about a factor of seven smaller and for the gate opposite the dot the lever arm is fourteen times smaller. From Coulomb blockade diamond measurements we find a typical Coulomb charging energy of 
1~meV, which corresponds roughly to the area of a single segment  in a disc-capacitor model.

\begin{figure}[t]
\centering
\includegraphics[width=7.5cm]{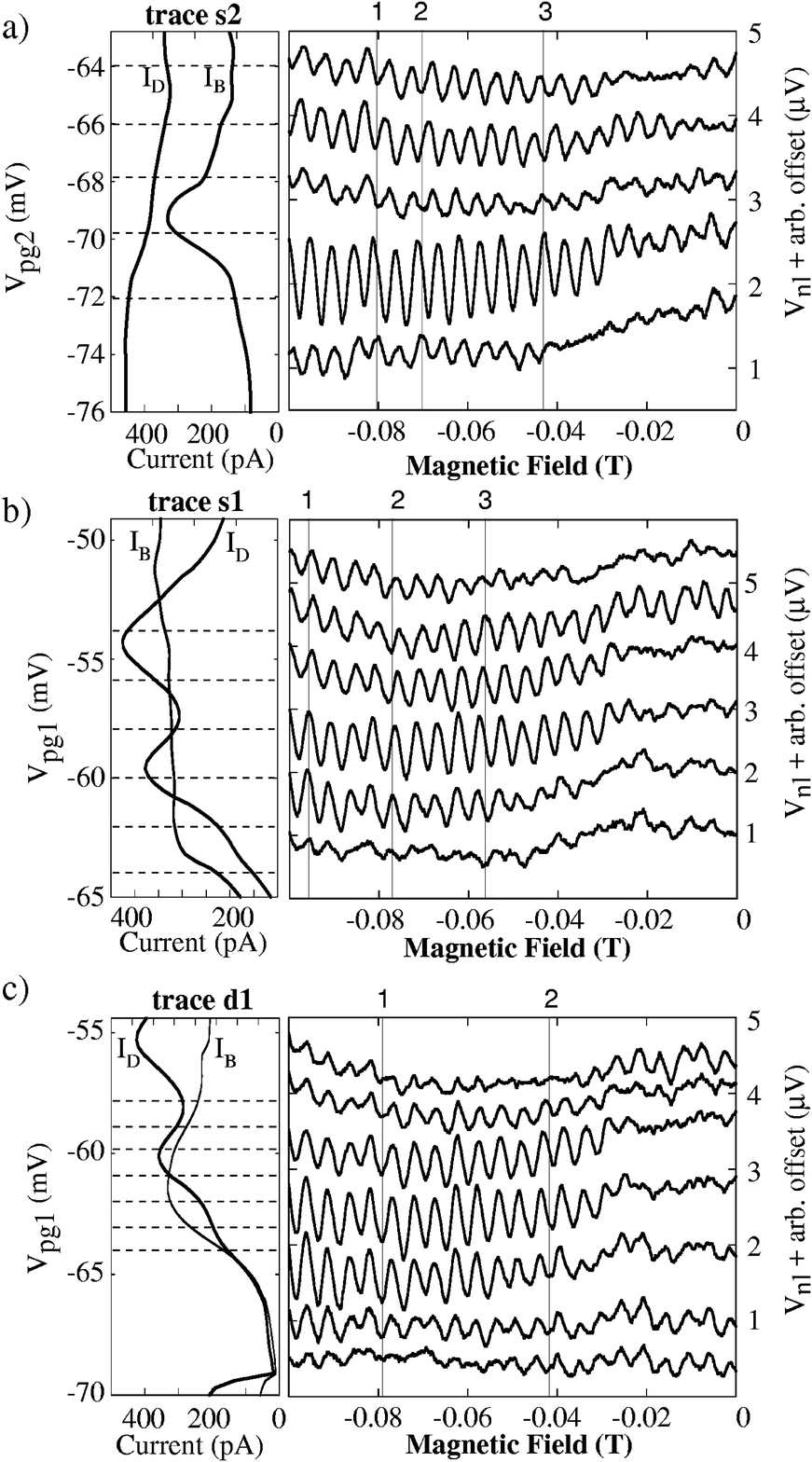}
\caption{\label{fig3} Currents $I_{\mathrm B}$ and $I_{\mathrm D}$ and AB-oscillations in the non-local voltage $V_\mathrm{nl}$ as a function of plunger gate voltages tuned along traces s2 (a), s1 (b) and d1 (c). The AB-oscillations are taken at the gate values indicated by the dashed lines. The thin vertical lines are shown only as a guide to the eye in order to visualize the phase shift more clearly.}
\end{figure}
In the following experiments we choose a regime where arm 1 and 2 are in the Coulomb blockade regime while the other two arms are left open and contain se\-ve\-ral lateral modes. We study a region in the plane defined by $V_{\mathrm{pg1}}$ and $V_{\mathrm{pg2}}$ close to a setting where both dots are tuned to a conductance maximum. This is shown schematically in Fig.~\ref{fig1}(b) where the dashed lines mark the Coulomb peak positions. The numbers in bra\-ckets indicate the electron occupation of the two dots. In our measurements we first keep one dot on a conductance maximum while stepping through a conductance peak in the other dot (traces s1 and s2 indicated by the arrows).  At the same time we detect the AB-oscillations in the non-local voltage for each value of the gate voltage. Figure~\ref{fig3}(a) shows an example where dot 2 is tuned along trace s2. The current signals on the left of the figure show a peak in the current $I_{\mbox{\tiny B}}$, i.e. the current in the terminal closer to the tuned dot. The current $I_{\mbox{\tiny D}}$ shows a weak monotonous dependence on pg2, as expected. In the main part of the figure on the right, AB-oscillations in the non-local voltage as a function of magnetic field are shown for selected gate voltages (dashed lines in the left part of the figure). We find that for very small magnetic fields the amplitude of the AB-oscillations is suppressed and it is difficult to make a statement about a phase change between the curves. The numbered thin vertical lines are a guide to the eye and connect maxima in the uppermost trace with mi\-ni\-ma
in the lowest trace. This indicates a phase shift by about $\pi$ as expected for the phase accumulation over a single isolated resonance~\cite{97schuster,01hackenbroich}. While the shift is continuous and positive (from top to bottom) for line 1 we find a slightly smaller shift for line 2 over the whole range of sweeps. A shift by $\pi$ takes place over the middle three curves. A change in phase or frequency for different values of the magnetic field could mean that in a semiclassical picture different paths around the ring contribute to the interference signal as the magnetic field is tuned. This illustrates that from our data a magnetic field independent AB-phase cannot be unambiguously defined.

Along trace s1 [Fig.~\ref{fig3}(b)] dot 1 is tuned while dot 2 is kept on a conductance resonance. Here, current $I_{\mathrm B}$ shows the expected flat behavior, but we observe a double peak structure in current $I_{\mathrm D}$. The phase change has the opposite sign compared to trace s2. For the vertical line 3 we find almost the expected phase change of $\pi$. For lines 1 and 2 the phase change is, however, smaller. While the magnetic field dependence of the phase needs further investigation we do find that the relative change of the transmission phase in both dots is the same as an electron is added to them. This means that the interference patterns are shifted in opposite directions.
In order to verify this we show the results for a sweep along the diagonal trace d1 where two electrons are added simultaneously to both dots [Fig~\ref{fig3}(c)]. 
For magnetic fields above 50~mT we find that the phase change over the conductance resonance is indeed $0=\pi-\pi$ as expected (see vertical line 1). However, we again find that deviations from this behavior occur for small magnetic fields where we find a small shift of the phase (vertical line 2). 

The design and fabrication process of our samples is different from the experiment by Schuster et al.~\cite{97schuster}.
In contrast to their experiment we have a second dot in the reference arm of the interferometer which gives us additional control of the relative phase change. 
Our fin\-dings indicate that the shift of the transmission phase of both dots has the same sign when they are tuned over a conductance resonance and the electron occupation is increased. In agreement with our expectation, we find a negligible shift for trace d1 where the phase difference between the arms of the ring remains constant. 

In conclusion, we have demonstrated that the phase evolution can be measured with individual control over the
electron occupancy in each dot. While the gene\-ral features are understood qualitatively, further work is needed to explain magnetic field dependent deviations from the theoretically expected phase changes. 
 \hyphenation{Ensslin}
\bibliographystyle{apsrev}
\bibliography{ab_phase}

\begin{thebibliography}{7}
\expandafter\ifx\csname natexlab\endcsname\relax\def\natexlab#1{#1}\fi
\expandafter\ifx\csname bibnamefont\endcsname\relax
  \def\bibnamefont#1{#1}\fi
\expandafter\ifx\csname bibfnamefont\endcsname\relax
  \def\bibfnamefont#1{#1}\fi
\expandafter\ifx\csname citenamefont\endcsname\relax
  \def\citenamefont#1{#1}\fi
\expandafter\ifx\csname url\endcsname\relax
  \def\url#1{\texttt{#1}}\fi
\expandafter\ifx\csname urlprefix\endcsname\relax\def\urlprefix{URL }\fi
\providecommand{\bibinfo}[2]{#2}
\providecommand{\eprint}[2][]{\url{#2}}

\bibitem[{\citenamefont{{M. B\"uttiker}}(1986)}]{86Bbuttiker}
\bibinfo{author}{\bibnamefont{{M. B\"uttiker}}}, \bibinfo{journal}{Phys. Rev.
  Lett.} \textbf{\bibinfo{volume}{57}}, \bibinfo{pages}{1761}
  (\bibinfo{year}{1986}).

\bibitem[{\citenamefont{Schuster et~al.}(1997)\citenamefont{Schuster, Buks,
  Heiblum, Mahalu, and Shtrikman}}]{97schuster}
\bibinfo{author}{\bibfnamefont{R.}~\bibnamefont{Schuster}},
  \bibinfo{author}{\bibfnamefont{E.}~\bibnamefont{Buks}},
  \bibinfo{author}{\bibfnamefont{M.}~\bibnamefont{Heiblum}},
  \bibinfo{author}{\bibfnamefont{D.}~\bibnamefont{Mahalu}}, \bibnamefont{and}
  \bibinfo{author}{\bibfnamefont{V.~U.~H.} \bibnamefont{Shtrikman}},
  \bibinfo{journal}{Nature} \textbf{\bibinfo{volume}{385}},
  \bibinfo{pages}{417} (\bibinfo{year}{1997}).

\bibitem[{\citenamefont{Buks et~al.}(1998)\citenamefont{Buks, Schuster,
  Heiblum, Mahalu, and Umansky}}]{98buks}
\bibinfo{author}{\bibfnamefont{E.}~\bibnamefont{Buks}},
  \bibinfo{author}{\bibfnamefont{R.}~\bibnamefont{Schuster}},
  \bibinfo{author}{\bibfnamefont{M.}~\bibnamefont{Heiblum}},
  \bibinfo{author}{\bibfnamefont{D.}~\bibnamefont{Mahalu}}, \bibnamefont{and}
  \bibinfo{author}{\bibfnamefont{V.}~\bibnamefont{Umansky}},
  \bibinfo{journal}{Nature} \textbf{\bibinfo{volume}{391}},
  \bibinfo{pages}{871} (\bibinfo{year}{1998}).

\bibitem[{\citenamefont{Ji et~al.}(2000)\citenamefont{Ji, Heiblum, Sprinzak,
  Mahalu, and Shtrikman}}]{00ji}
\bibinfo{author}{\bibfnamefont{Y.}~\bibnamefont{Ji}},
  \bibinfo{author}{\bibfnamefont{M.}~\bibnamefont{Heiblum}},
  \bibinfo{author}{\bibfnamefont{D.}~\bibnamefont{Sprinzak}},
  \bibinfo{author}{\bibfnamefont{D.}~\bibnamefont{Mahalu}}, \bibnamefont{and}
  \bibinfo{author}{\bibfnamefont{H.}~\bibnamefont{Shtrikman}},
  \bibinfo{journal}{Science} \textbf{\bibinfo{volume}{290}},
  \bibinfo{pages}{779} (\bibinfo{year}{2000}).

\bibitem[{\citenamefont{Ji et~al.}(2002)\citenamefont{Ji, Heiblum, and
  Shtrikman}}]{02ji}
\bibinfo{author}{\bibfnamefont{Y.}~\bibnamefont{Ji}},
  \bibinfo{author}{\bibfnamefont{M.}~\bibnamefont{Heiblum}}, \bibnamefont{and}
  \bibinfo{author}{\bibfnamefont{H.}~\bibnamefont{Shtrikman}},
  \bibinfo{journal}{Phys. Rev. Lett.} \textbf{\bibinfo{volume}{88}},
  \bibinfo{pages}{076601} (\bibinfo{year}{2002}).

\bibitem[{\citenamefont{Hackenbroich}(2001)}]{01hackenbroich}
\bibinfo{author}{\bibfnamefont{G.}~\bibnamefont{Hackenbroich}},
  \bibinfo{journal}{Phys. Rep.} \textbf{\bibinfo{volume}{343}},
  \bibinfo{pages}{463} (\bibinfo{year}{2001}).

\bibitem[{\citenamefont{Fuhrer et~al.}(2002)\citenamefont{Fuhrer, Dorn,
  L{\"u}scher, Heinzel, Ensslin, Wegscheider, and Bichler}}]{02fuhrer}
\bibinfo{author}{\bibfnamefont{A.}~\bibnamefont{Fuhrer}},
  \bibinfo{author}{\bibfnamefont{A.}~\bibnamefont{Dorn}},
  \bibinfo{author}{\bibfnamefont{S.}~\bibnamefont{L{\"u}scher}},
  \bibinfo{author}{\bibfnamefont{T.}~\bibnamefont{Heinzel}},
  \bibinfo{author}{\bibfnamefont{K.}~\bibnamefont{Ensslin}},
  \bibinfo{author}{\bibfnamefont{W.}~\bibnamefont{Wegscheider}},
  \bibnamefont{and} \bibinfo{author}{\bibfnamefont{M.}~\bibnamefont{Bichler}},
  \bibinfo{journal}{Superlattice Microstruct.} \textbf{\bibinfo{volume}{31}},
  \bibinfo{pages}{19} (\bibinfo{year}{2002}).

\end{thebibliography}

\end{document}